\documentclass[particles,article,accept,pdftex,moreauthors]{Definitions/mdpi} 
\nolinenumbers
\usepackage{subfig}
\usepackage{soul}
\firstpage{1} 
\pubvolume{8}
\issuenum{3}
\articlenumber{69}
\pubyear{2025}
\copyrightyear{2025}
\externaleditor{Matteo Duranti and Valerio Vagelli} 
\datereceived{17 June 2025} 
\daterevised{10 July 2025} 
\dateaccepted{16 July 2025} 
\datepublished{19 July 2025} 
\hreflink{https://\\doi.org/10.3390/particles8030069} 

\Title{Background Measurements and Simulations of the ComPair Balloon Flight}

\TitleCitation{Background Measurements and Simulations of the ComPair Balloon Flight}

\Author{Zachary Metzler 
 $^{1,2,3}$*\orcidA{}, Nicholas Kirschner $^{2,4}$, Lucas Smith $^{1,2,3}$, Nicholas Cannady $^{2}$\orcidB{}, Makoto Sasaki $^{1,2,3}$, \linebreak  Daniel Shy $^{5}$, Regina Caputo $^{2}$, Carolyn Kierans $^{2}$\orcidC{}, Aleksey Bolotnikov $^{6}$\orcidD{}, Thomas J. Caligiure $^{7}$, \mbox{Gabriella A. Carini $^{6}$}, Alexander 
 Wilder Crosier $^{7}$, Jack Fried $^{6}$, Priyarshini Ghosh $^{2,3,8}$, Sean Griffin $^{9}$\orcidE{}, \mbox{Jon Eric Grove $^{5}$}, \mbox{Elizabeth Hays $^{2}$},   Sven Herrmann $^{6}$, Emily Kong $^{10}$, Iker Liceaga-Indart $^{2}$, Julie McEnery $^{2}$, \mbox{John Mitchell $^{2}$}, Alexander A. Moiseev $^{1,2,3}$,   Lucas Parker $^{11}$, Jeremy Perkins $^{2}$, Bernard Phlips $^{5}$, \mbox{Adam J. Schoenwald $^{2}$}, Clio Sleator $^{5}$, David J. Thompson $^{2}$\orcidF{},   Janeth Valverde $^{2,3,8,12}$, Sambid Wasti $^{2,3,13}$\orcidG{}, \mbox{Richard Woolf $^{5}$\orcidH{}}, Eric Wulf $^{5}$ \mbox{and Anna Zajczyk $^{2,3,8}$}}

\AuthorNames{Zachary Metzler, Nicholas Kirschner, Lucas Smith, Nicholas Cannady, Makoto Sasaki, Daniel Shy, Regina Caputo, Carolyn Kierans, Aleksey Bolotnikov, Thomas J. Caligiure, Gabriella A. Carini, A. Wilder Crosier, Jack Fried, Priyarshini Ghosh, Sean Griffin, J. Eric Grove, Elizabeth Hays, Sven Herrmann, Emily Kong, Iker Liceaga-Indart, Julie McEnery, John Mitchell, Alexander A. Moiseev, Lucas Parker, Jeremy Perkins, Bernard Phlips, Adam J. Schoenwald, Clio Sleator, David J. Thompson, Janeth Valverde, Sambid Wasti, Richard Woolf, Eric Wulf, and Anna Zajczyk}

\isAPAStyle{%
       \AuthorCitation{Lastname, F., Lastname, F., \& Lastname, F.}
         }{%
        \isChicagoStyle{%
        \AuthorCitation{Lastname, Firstname, Firstname Lastname, and Firstname Lastname.}
        }{
        \AuthorCitation{Metzler, Z.; Kirschner, N.; Smith, L.; Cannady, N.; Sasaki, M.; Shy, D.; Caputo, R.; Kierans, C.; Bolotnikov, A.; Caligiure, T.J.; et al.}}
        }

\address{%
$^{1}$ \quad Department of Physics, University 
 of Maryland, College Park
, MD 20740, USA
; ldsmith@umd.edu (L.S.); msasaki@umd.edu (M.S.); amoiseev@umd.edu (A.A.M.) 
\\

$^{2}$ \quad NASA Goddard Space Flight Center, Greenbelt, MD 20771, USA
; nicholas.j.kirschner@nasa.gov (N.K.); nick.cannady@gmail.com (N.C.); regina.caputo@nasa.gov (R.C.);  carolyn.a.kierans@nasa.gov (C.K.); priyarshini.ghosh@nasa.gov (P.G.); elizabeth.a.hays@nasa.gov (E.H.); iker.liceagaindart@nasa.gov (I.L.-I.); jmcenery@umd.edu (J.M.); john.w.mitchell@nasa.gov (J.M.); jeremy.s.perkins@nasa.gov (J.P.); adam.schoenwald@nasa.gov (A.J.S.); david.j.thompson@nasa.gov (D.J.T.); janeth.valverde@marquette.edu~(J.V.); sambid.wasti@gmail.com (S.W.);  anna.zajczyk@gmail.com (A.Z.)\\

$^{3}$ \quad Center for Research and Exploration in Space Science \& Technology II, Greenbelt, MD 20771, USA 
\\

$^{4}$ \quad Department of Physics, George Washington University, Washington, DC 20052, USA 
\\

$^{5}$ \quad U.S. Naval Research Laboratory, Washington, DC 20375, USA
; daniel.shy.civ@us.navy.mil (D.S.); eric.grove@nrl.navy.mil (J.E.G.); bernard.f.phlips.civ@us.navy.mil (B.P.); clio.c.sleator.civ@us.navy.mil (C.S.); \linebreak  richard.woolf@nrl.navy.mil (R.W.); wulf@nrl.navy.mil (E.W.)\\

$^{6}$ \quad Brookhaven National Laboratory, Upton, NY 11973, USA
; bolotnik@bnl.gov (A.B.); carini@bnl.gov (G.A.C.); jfried@bnl.gov (J.F.); sherrmann@bnl.gov (S.H.) \\

$^{7}$ \quad {Naval Research Enterprise Internship Program, U.S. Naval Research Laboratory}, Washington, DC 20375, USA
; tcaligiure@ufl.edu (T.J.C.); alexanderwcrosier@gmail.com (A.W.C.)\\

$^{8}$ \quad Department of Physics, University 
 of Maryland, Baltimore County, MD 21250, USA
\\

$^{9}$ \quad Wisconsin IceCube Particle Astrophysics Center, Madison, WI 53706, USA
; sean.griffin@icecube.wisc.edu  \\

$^{10}$\quad Technology Service Corporation, Arlington, VA 22202, USA
; emily.kong.98@gmail.com \\

$^{11}$\quad Los Alamos National Laboratory, Los Alamos, NM 87545, USA
; lpp@lanl.gov  \\

$^{12}$\quad Department of Physics, Marquette University, Milwaukee, WI 53201, USA \\

$^{13}$\quad Department of Physics, Catholic University 
 of America, Washington, DC 20064, USA 
}

\corres{Correspondence: \textcolor{black}{zmetzler@umd.edu}}

\abstract{ComPair, a prototype of the All-sky Medium Energy Gamma-ray Observatory (AMEGO), completed a short-duration high-altitude balloon campaign on 27 August 2023 
 from Fort Sumner, New Mexico, USA. The goal of the balloon flight was to demonstrate ComPair as both a Compton and Pair telescope in flight, reject the charged particle background, and measure the background $\gamma$-ray spectrum. This analysis compares measurements from the balloon flight with Monte Carlo simulations to benchmark the instrument. The comparison finds good agreement between the measurements and simulations and supports the conclusion that ComPair accomplished its goals for the balloon campaign. Additionally, two charged particle background rejection schemes are discussed: a soft ACD veto that records a higher charged particle event rate but with less risk of event loss, and a hard ACD veto that limits the charged particle event rate on board. There was little difference in the measured spectra from the soft and hard ACD veto schemes, indicating that the hard ACD veto could be used for future flights. The successes of ComPair's engineering flight will inform the development of the next generation of ComPair with upgraded detector technology and larger active area.}

\keyword{gamma ray; ComPair; MeV $\gamma$-rays; astroparticle; instrumentation; AMEGO; balloon} 

\begin{document}

\section{Introduction}\label{sec:Intro}

\begin{figure}[H]
    \includegraphics[width=0.75\linewidth]{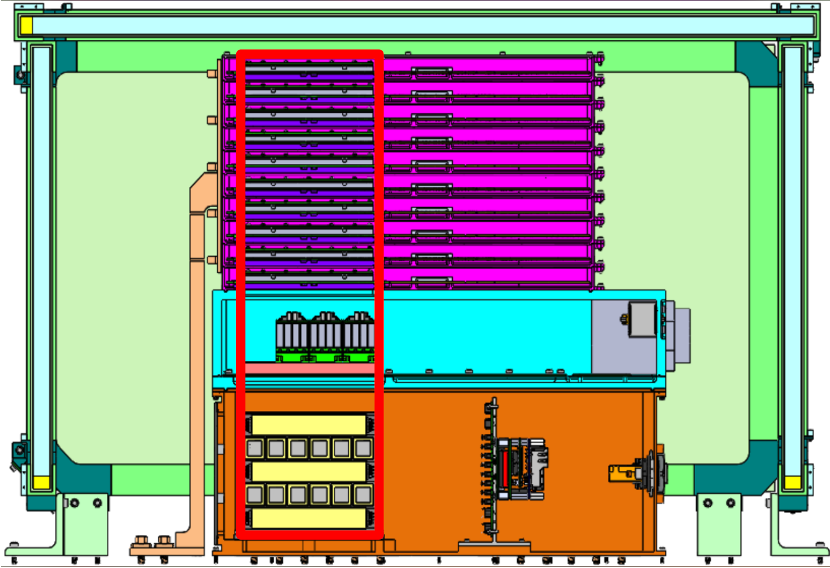}
    \caption{\textcolor{black}{ComPair consists of four detector subsystems: a DSSD Tracker (purple), a~CZT Calorimeter (blue), a~CsI Calorimeter (orange), and~a plastic scintillator ACD (green). The~red box highlights the detector stack, which contains the active area of the Tracker and Calorimeters.}}
    \label{fig:ComPair}
\end{figure}

The lack of sensitivity to astrophysical photons between $\sim$100 keV and $\sim$100 MeV relative to energies immediately above and below has earned this energy range the moniker ``MeV Gap.'' There are multiple factors contributing to the MeV gap, including (a) the need to discriminate between an image with both Compton scattering and pair production events, (b) a minimum in the total photon interaction cross section, and~(c) a dominant charged particle background. 
The All-sky Medium Energy Gamma-ray Observatory (AMEGO) \cite{AMEGO_McEnery_2019,AMEGO_Kierans_2020} is a NASA probe-class mission concept designed to overcome the challenges in order to fill the MeV gap.
ComPair~\cite{ComPair_Shy_2022,ComPair_Valverde_2023} is a prototype of AMEGO and  consists of four detector subsystems: a 10-layer double-sided silicon strip detector Tracker~\cite{tracker_griffin_2019, tracker_griffin_2020, tracker_kirschner_2024}, a~Cadmium Zinc Telluride (CZT) Calorimeter~\cite{CZT_Hays_2019}, a~thallium-doped Cesium Iodide (CsI) Calorimeter~\cite{CsI_Woolf_2019, CsI_Flight_Shy_2024}, and~a plastic scintillator Anti-Coincidence Detector (ACD) \cite{ACD_Metzler_2024}. \textcolor{black}{A diagram of ComPair is shown in Figure~\ref{fig:ComPair}.} The four detector systems work together to detect and characterize $\gamma$-rays. At~low energies, the~tracker serves as the Compton scatterer, and~at high energies, it acts as the pair conversion material. The~CZT Calorimeter enhances the low energy sensitivity with good energy and position resolution, while the CsI Calorimeter enhances the high energy sensitivity with its stopping power. The~ACD serves as an active shield to suppress the charged particle background, because~it is generally insensitive to $\gamma$-rays but efficiently detects charged~particles.

\textcolor{black}{On 27 August 2023
, ComPair} conducted a \textcolor{black}{6 h, 17 min} high-altitude balloon flight from Fort Sumner, New Mexico, USA. The~goals of this engineering flight \textcolor{black}{were} to demonstrate that the integrated system works as a Compton and Pair telescope in a space-like environment, to~verify that the charged particle background rejection operates as intended, and~to measure the background $\gamma$-ray spectrum. The successful completion of a short-duration flight is also a prerequisite for subsequent long-duration balloon~flights.

Earlier papers have reported on the ComPair balloon flight. 
Kirschner~et~al., 2024~\cite{tracker_kirschner_2024}, Shy~et~al., 2024~\cite{CsI_Flight_Shy_2024}, and~Metzler~et~al., 2024~\cite{ACD_Metzler_2024} each focus individually on the Tracker, CsI Calorimeter, and~ACD, respectively.
Smith~et~al. 2024~\cite{ComPair_Flight_Smith_2024} highlights much of the pre-flight validation that is not considered here and also presents the measured flight spectra for each subsystem without applying the ACD veto or \textcolor{black}{performing} event reconstruction.
In this analysis, we will demonstrate that the ComPair engineering flight achieved its goals by validating the measured ACD veto rate and event reconstruction with Monte Carlo simulations using MEGAlib's Cosima tool~\cite{MEGAlib}. Section~\ref{sec:Methods} describes the data analysis pipeline, simulations, and~balloon flight in general. Sections~\ref{sec:SoftVeto} and \ref{sec:reconstruction} compare the measured data to the background simulations, and~Section~\ref{sec:HardVeto} compares the effectiveness of a hard ACD veto, in~which events are rejected during data collection, to~a soft ACD veto, in~which events are rejected only during data analysis, to~inform their use on future ComPair balloon flights and~AMEGO.

\section{Materials and~Methods}\label{sec:Methods}
\unskip

\subsection{Event~Selection}

When a $\gamma$-ray or charged particle interacts in one of the detector subsystems, the~subsystem will send a signal to the Trigger Module~\cite{TM_Sasaki_2020}. We call these signals `primitives'. An event is defined by a unique event ID, which is created by the Trigger Module when primitives matching any of the following \textcolor{black}{combinations} are received from the detector subsystems \textcolor{black}{within 20 $\upmu$s}.

\begin{enumerate}
    \item Any 2 Tracker primitives
    \item A single Tracker primitive and the CZT primitive
    \item A single Tracker primitive and the CsI primitive
    \item The CZT primitive and the CsI primitive
\end{enumerate}

The Tracker produces one primitive for each side of each layer, and the CZT, CsI, and~ACD each produce one primitive for the entire~subsystem. 

The \textcolor{black}{event IDs} serve two purposes: event alignment and noise suppression. Each detector system has its own calibration pipeline, and~the unique event ID is used to match simultaneous interactions in multiple detector systems. Requiring multiple primitives to define an event ID also suppresses noise \textcolor{black}{that could falsely trigger a single channel.} 

The Trigger Module also defines the two ACD veto modes, namely soft and hard. During~soft ACD veto, the~Trigger Module provides an event ID whether or not the ACD produces a primitive. During~hard ACD veto, event IDs are provided only when the ACD does not produce a~primitive.

\subsection{Simulations and Event~Reconstruction}

\begin{figure}[H]
    \includegraphics[width=1.0\linewidth]{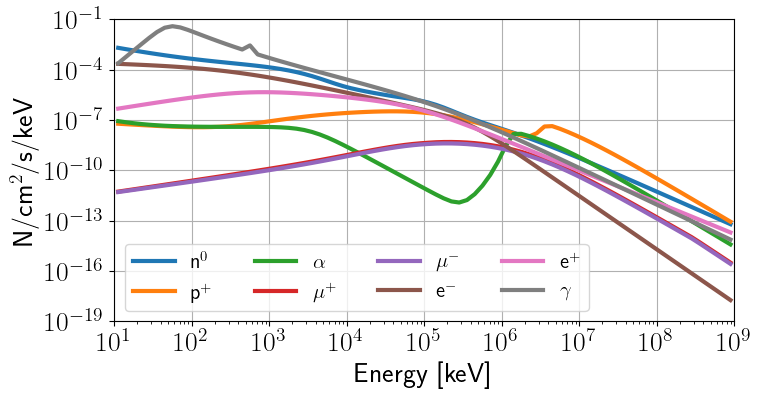}
    \caption{The angle-integrated particle background spectra for the ComPair balloon flight with \mbox{8 particle} species as calculated with EXPACS~\cite{EXPACS}. EXPACS also provides azimuth-dependent spectra for each of these species, which were used to simulate the conditions during the flight with MEGAlib's Cosima~\cite{MEGAlib}. The~settings used to produce this spectra are 34$^\circ$ latitude, $-$104$^\circ$ longitude, 40 km above sea level, 124 for the W value which \textcolor{black}{is} a measure of solar activity \textcolor{black}{approaching solar maximum}, and~an ideal atmosphere surrounding~environment.}
    \label{fig:EXPACS}
\end{figure}

To benchmark the measurements, we performed simulations of the high-altitude particle background using the EXcel-based Program for calculating the Atmospheric Cosmic-ray Spectrum (EXPACS) \cite{EXPACS}. EXPACS uses longitude, latitude, altitude, and~solar activity to produce energy- and angular-dependent spectra for $\gamma$-rays, $\text{e}^+$, $\text{e}^-$, $\text{p}^+$, $\text{n}^0$, $\mu^+$ and $\mu^-$, and~$\alpha$ particles. Figure~\ref{fig:EXPACS} shows the angle-integrated spectra for each particle species. A~Monte Carlo simulation is performed with MEGAlib's GEANT4-based Cosima \mbox{tool~\cite{MEGAlib,GEANT4}}. The~Cosima simulation \textcolor{black}{uses a detailed mass model of the ComPair instrument and a simplified model of the gondola. It} records any interaction from $\gamma$-rays and particles in the active material of any subsystem, including the~ACD. 

The simulation includes tracking delayed emission due to nuclear activation by neutrons. The~activation simulation is a three-step process. (1) An initial exposure tracks leptonic, photonic, and~prompt emission due to nuclear excitations. This first phase also records the production of excited nuclei with longer decay times. (2) A build-up phase quickly extrapolates the production of the longer-lived nuclear excitations at the same rate as \textcolor{black}{in} step (1). (3) The last phase propagates photons produced by the delayed emission of the longer-lived nuclear excitations. The~times used for the ComPair balloon flight activation simulation were a 15-min exposure, 1-h activation build-up, and~15 min tracking the delayed nuclear~de-excitations.

The simulated events are passed through the ComPair Detector Effects Engine (DEE), an~internally developed python pipeline that applies instrumental effects and transforms the simulated interactions into a form resembling the raw output from the detectors. The~instrumental effects include chance coincidence between multiple incident particles, detector dead time, trigger logic, thresholding, convolutions with noise pedestals, and~converting to ADC values. Finally, the~output of the DEE is passed through the same data analysis pipeline as the measured~data.

\begin{figure}[H]
    \includegraphics[width=1.0\linewidth]{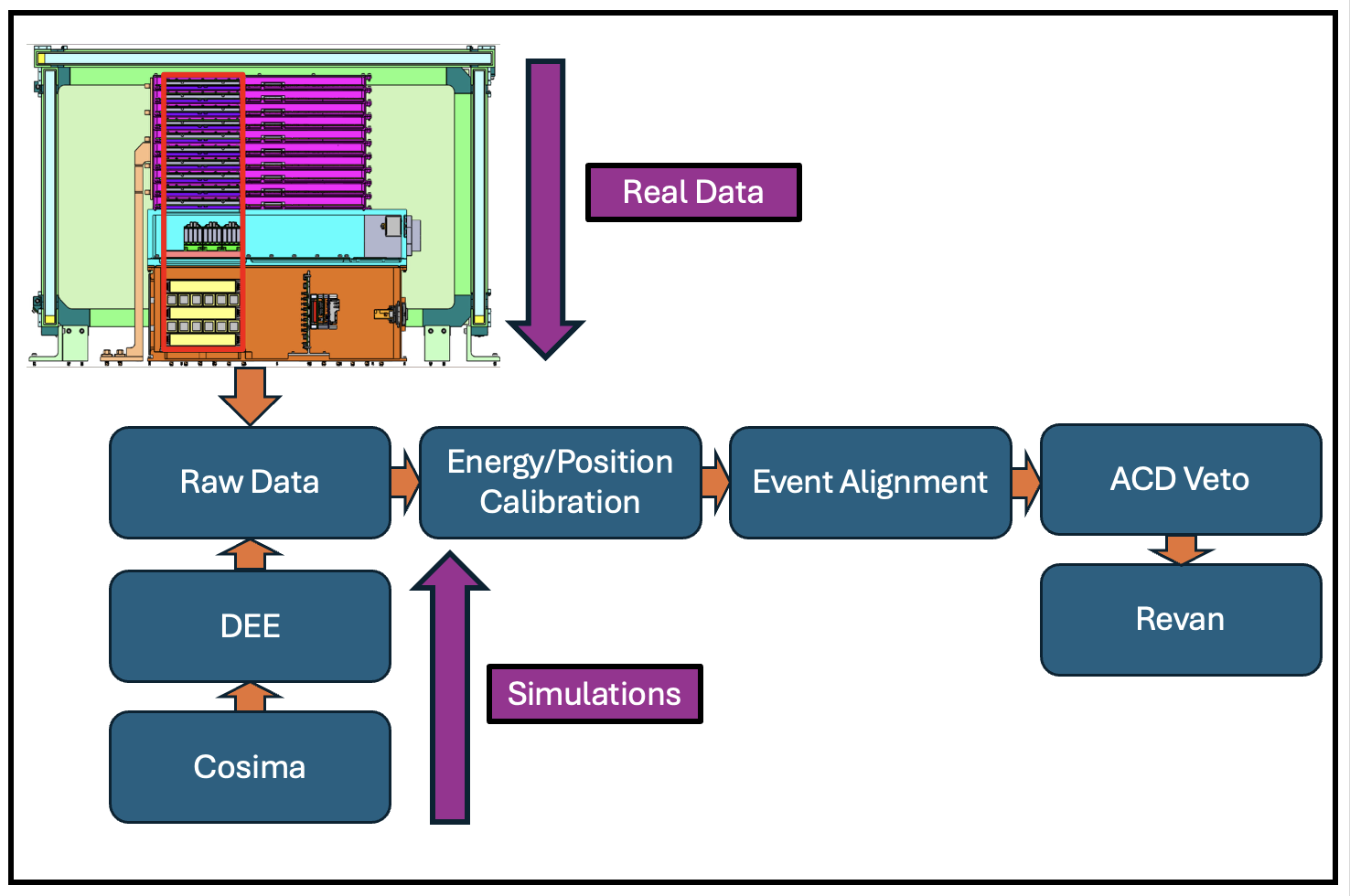}
    \caption{A schematic showing the steps performed on real and simulated data. Simulated interactions are produced with MEGAlib's Cosima~\cite{MEGAlib} and transformed by the DEE to resemble raw data. Real raw data comes directly from the detector subsystems. Both real and simulated raw data undergoes energy and position calibrations, event alignment, an~ACD veto, and~event reconstruction offline. Event reconstruction classifies the remaining events into PH, CO, PA, and~MU~events.}
    \label{fig:DEE1}
\end{figure}

Figure~\ref{fig:DEE1} shows a simple schematic of the steps for processing real and simulated data. The~data analysis pipeline includes energy and position calibrations, event alignment between the detector subsystems \textcolor{black}{using the event IDs produced by the TM during data collection, and} vetoing events that triggered the~ACD.

\begin{figure}[H]
    \includegraphics[width=1.0\linewidth]{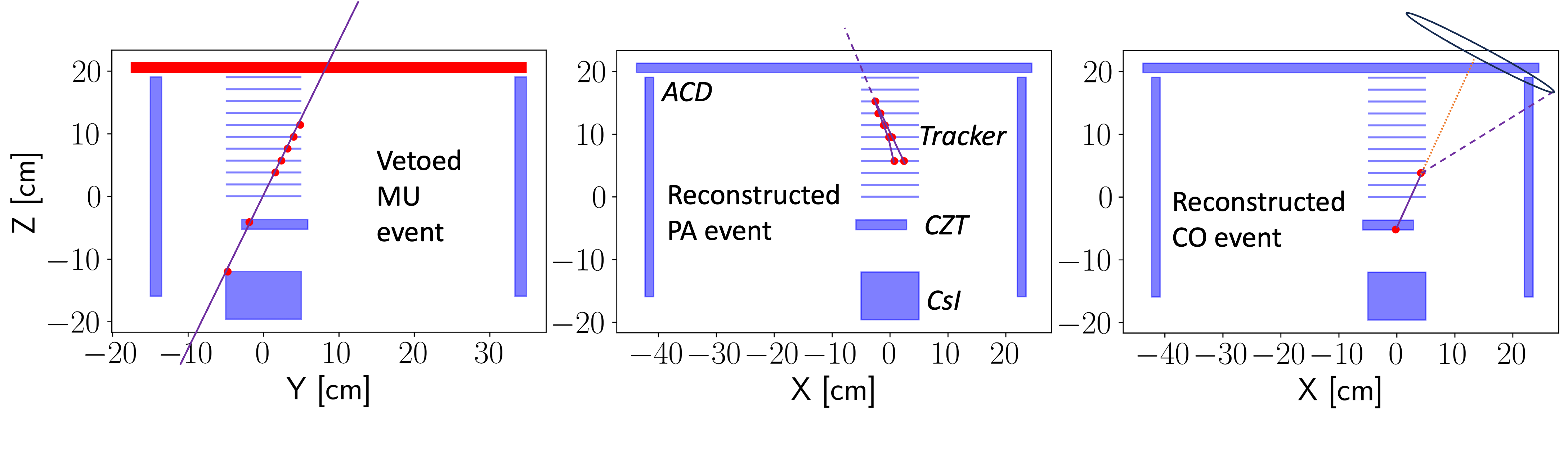}
    \caption{Examples of MU, PA, and~CO events from the balloon flight, where the triggered hits are shown as red dots. A~MU event triggers the top ACD panel, consists of a single track, and~is not used for imaging. A~PA event consists of two tracks that intersect at a vertex, and~can be pointed back to its source along a single direction. This PA event seems to stop in the Tracker, because~the tracks exit the side of the Tracker before reaching the CZT Calorimeter. Any events with multiple hits that are not MU or PA are reconstructed as CO events if a valid Compton scattering sequence can be found. Events with a valid Compton scattering sequence can point back to a ring in the sky where the photon originated, events without a valid sequence are~rejected.}
    \label{fig:EventDisplay}
\end{figure}

Event reconstruction is performed with MEGAlib's Revan tool~\cite{MEGAlib}, which classifies the remaining events as photoabsorption (PH), Compton scattering (CO), pair production (PA), or~muon tracks (MU), although~particle MU events that pass the ACD veto are predominantly $n^0~\mathrm{and} ~p^+$ at float altitude. Figure~\ref{fig:EventDisplay} shows examples of CO, PA, and~MU events from real flight data. Among~multi-hit events, Revan will first search for PA events (center panel of Figure~\ref{fig:EventDisplay}), which are fit by two tracks that meet at a vertex. If~the event is not a PA event, Revan will look for MU events, which are defined by a single track. An~example vetoed MU event is shown in the left panel of Figure~\ref{fig:EventDisplay}. \textcolor{black}{Reconstructed PA events point directly to the $\gamma$-ray source, while MU events are not used for imaging.} Finally, Revan will attempt to find a valid Compton scattering sequence for any non-PA and non-MU events. If~a valid sequence is found, the~event is classified as a CO event (right panel of Figure~\ref{fig:EventDisplay}). Reconstructed CO events project to a ring in the sky that overlaps with the $\gamma$-ray source. Any event without a valid Compton scattering sequence is~rejected.

\unskip


\unskip

\subsection{Balloon Flight~Overview}

\begin{figure}[H]
    \centering
    \includegraphics[width=1.0\linewidth]{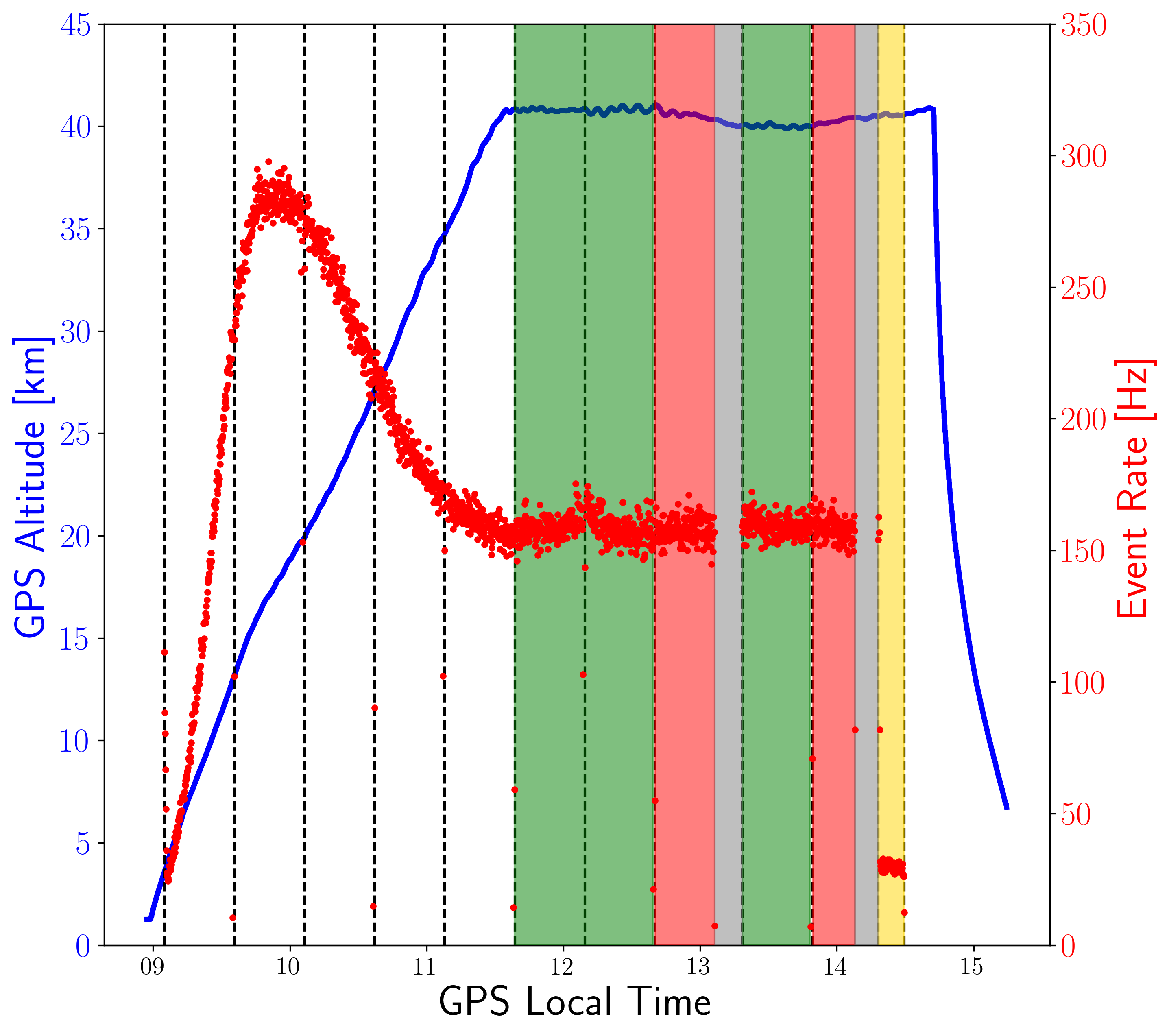}
    \caption{Blue: the altitude throughout the balloon flight. Red: the number of unique event IDs produced per second during the flight. The~times given are the local date and time in New Mexico, USA. The~green bands indicate the times used to analyze the soft ACD veto, and~the yellow band covers the hard ACD veto. \textcolor{black}{The gray segments are gaps in the data due to overheating of the main power distribution unit, which had to be turned off and allowed to cool before restarting operations. The~red bands show times, for~which Tracker data was unrecoverable due to the abrupt shut-down in response to the thermal issues.}}
    \label{fig:rates_and_altitude}
\end{figure}

ComPair's balloon flight from Ft Sumner, NM, USA in 2023 was 6 h and 17 min long, of~which 3 h and 13 min were over 40 km above sea level.  Figure~\ref{fig:rates_and_altitude} shows the altitude and event rate as a function of time. Data were collected in 30-min segments during the flight, marked by the vertical dashed lines. The~green bands represent the soft ACD veto, and~the yellow band covers a test of the hard ACD veto. The~hard ACD veto substantially reduces the event rate but increases the risk of event loss if the ACD malfunctions. The~green bands total 90 min, and~the yellow band covers 11~min.

The DC-DC converter for the main power distribution unit's temperature became too hot to safely operate during the two gray segments, so we turned ComPair off to cool to an~appropriate temperature. Unfortunately, the~Tracker data was not recoverable for the two red segments, so only the times in the green bands of Figure~\ref{fig:rates_and_altitude} are used for the analysis of the soft veto mode at float altitude. \textcolor{black}{The GPS and TM data was still recoverable during the red segments, so the altitude and event rates for these times are still shown.} The two gray segments span 12 and 10 min, and~the red segments span 26 and 19 min. 

\unskip

\section{Results and~Discussion}\label{sec:Results}
\unskip

\subsection{Soft ACD Veto~Efficiency}\label{sec:SoftVeto}

A key benchmark for ComPair's balloon flight is the effectiveness of the ACD to reject charged particles while passing (not vetoing) $\gamma$-rays. Table~\ref{tab:SoftSimVetoRate} shows event rates before and after applying the soft ACD veto, wherein events are rejected after the event alignment step of Figure~\ref{fig:DEE1} if there is an interaction in the ACD. The~first two rows show the overall measured and simulated event rates, respectively. \textcolor{black}{The other rows show} the event rates for each simulated particle species, where activation is defined as the delayed emission due to nuclear de-excitations and coincidence is when any 2 simulated particles interact within 20~\textcolor{black}{$\upmu$s}. The~real measured data rate is $\approx$20\% greater than the simulated total. This could be due to the simulated geometry not accounting for all of the passive material aboard the flight gondola or other unknown effects not included in the DEE. Before~applying the ACD veto, the~dominant particle types are protons and alpha particles, while $\gamma$-rays are third at just under $15\%$ of the simulated events. After~applying the veto, $\gamma$-rays make up just over $68\%$ of the simulated events and now dominate over the charged particle background, validating the efficiency of the~ACD. 

\begin{figure}[H]
    \includegraphics[width=0.95\linewidth]{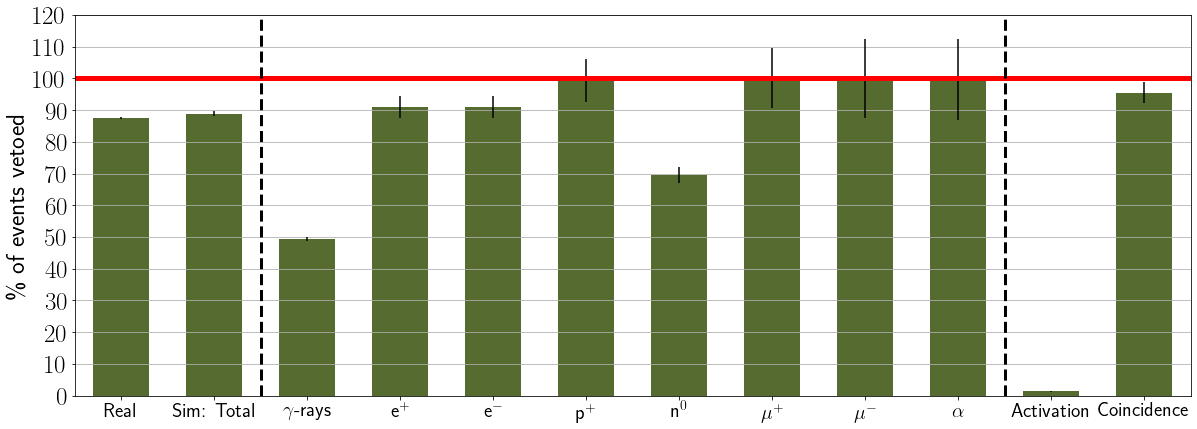}
    \caption{A comparison of the veto rates between the real and simulated data. The~red line at $100\%$ corresponds to all events being rejected. The~first column shows the measured data, the~second column shows the total of the simulated data, columns 3--10 show the simulation divided by particle type. Column 11 represents the simulated events due to activation within the instrument or passive material. Column 12 is for events where two simulated particles interacted with the detector within 20 $\upmu$s, and~so were assigned to a single~event.}
    \label{fig:SoftSimVetoPercent}
\end{figure}

Figure
~\ref{fig:SoftSimVetoPercent} shows the same information as Table~\ref{tab:SoftSimVetoRate} as the percentage of each particle type that is vetoed by the ACD. Over~$99\%$ of protons, muons, and~alpha particles are rejected by the ACD, while most $\gamma$-rays and events due to activation pass. Activation is particularly challenging to veto, because~the delayed de-excitation is not coincident with the incident particle. Therefore, the~primary particle that causes the excitation could be vetoed, but~the delayed emission likely will not \textcolor{black}{be vetoed}.

\begin{figure}[H]
    \includegraphics[width=0.8\linewidth]{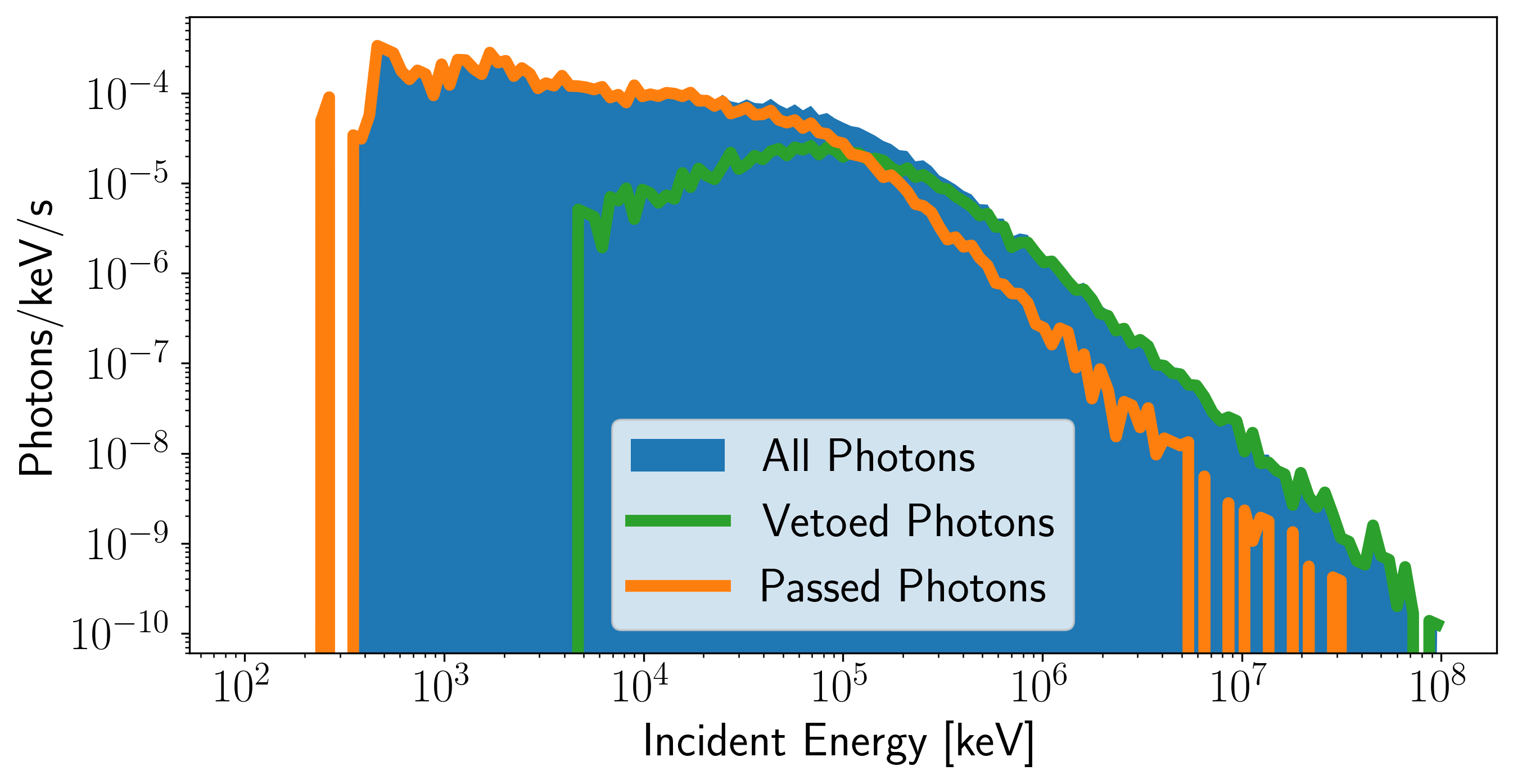}
    \caption{A spectrum of the incident energies of simulated photons that satisfied ComPair's trigger conditions. The~total spectrum is shown in blue, while the orange and green lines show the spectra for passed (not vetoed) and vetoed photons, respectively. This shows that the 50\% of $\gamma$-rays that are vetoed are high energy photons, which likely produce a shower that triggers the~ACD.}
    \label{fig:SimPhotonVeto}
\end{figure}

\begin{figure}[H]
    \includegraphics[width=0.75\linewidth]{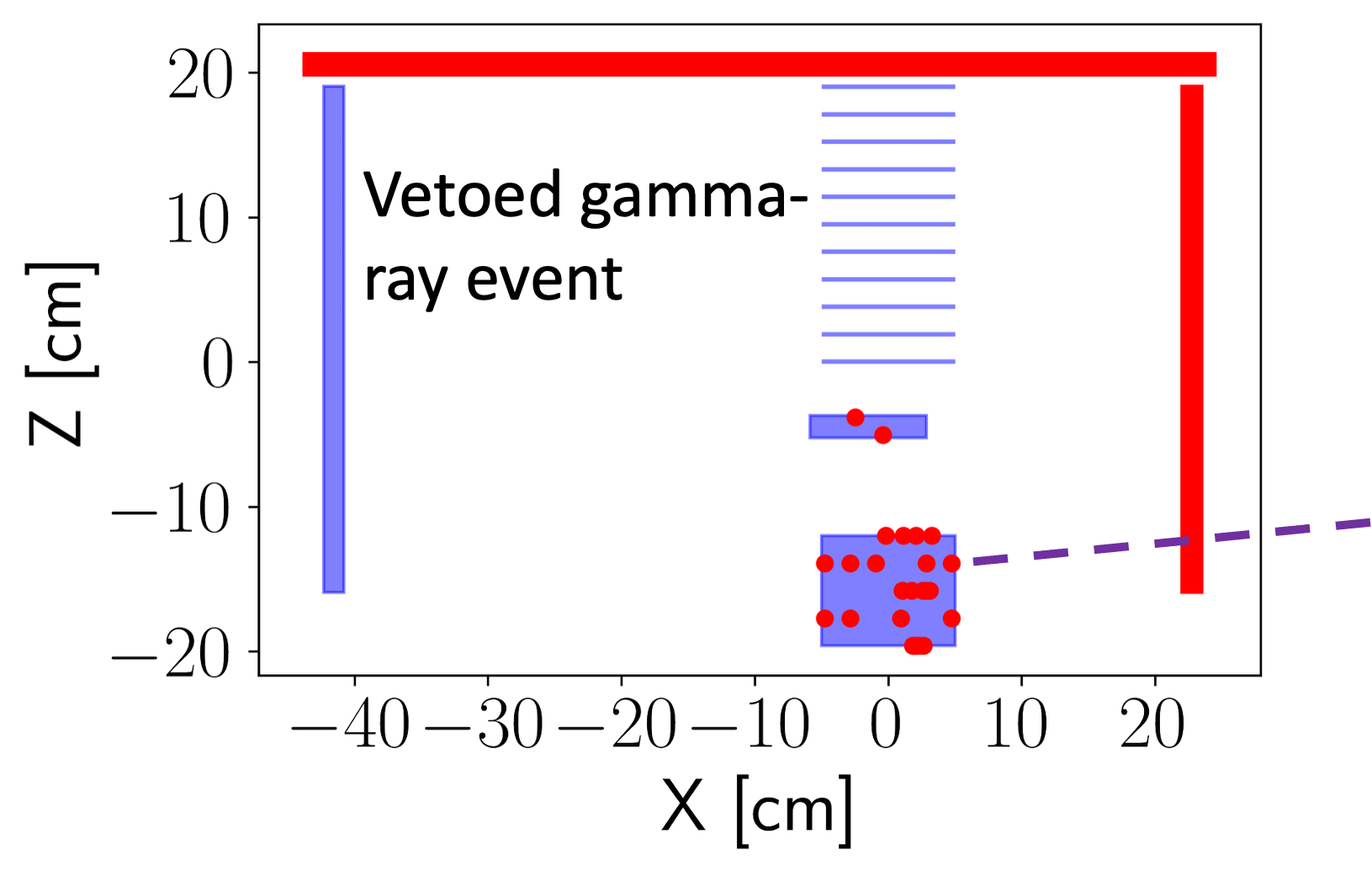}
    \caption{\textcolor{black}{An example event display of a simulated vetoed $\gamma$-ray with an incident energy of 29 GeV. The~$\gamma$-ray's initial direction is shown with a purple dashed line, and~the red dots indicate interaction locations, although~the order of interactions was not saved during the simulation.}}
    \label{fig:Veto_Photon_Event_Display}
\end{figure}

Figure~\ref{fig:SoftSimVetoPercent} shows that $\sim$50\% of $\gamma$-rays are vetoed by the ACD, and~it is important to understand the cause.
Figure~\ref{fig:SimPhotonVeto} shows the total spectrum of incident simulated $\gamma$-ray energies that satisfy ComPair's trigger conditions in blue as well as the spectra for passed and vetoed $\gamma$-rays in orange and green, respectively. There is a crossover between the passed and vetoed $\gamma$-ray spectrum around 100 MeV, above~which a majority of the incident $\gamma$-rays are vetoed by the ACD. \textcolor{black}{These vetoed $\gamma$-ray events predominantly pair produce in the Calorimeters or the aluminum base plate, on~which ComPair is mounted. The~high-energy $e^-$s and $e^+$s produced in the shower ultimately trigger an ACD panel, vetoing the event. Figure~\ref{fig:Veto_Photon_Event_Display} shows an event display for one of the vetoed $\gamma$-ray events with an incident energy of 29 GeV. The~purple dashed line shows the incident direction of the $\gamma$-ray, but~the order of interactions was not saved during the simulation.} A larger instrument, like AMEGO, would model the showers produced by high energy $\gamma$-rays and override the ACD veto for these types of events, but~this is beyond the scope for~ComPair.


\begin{table}[H] 
\caption{\textcolor{black}{A comparison of the event rates before and after the soft ACD veto is applied. The~table shows the totals of the measured and simulated data, the~simulation divided by particle type, simulated nuclear activation caused by high energy protons, neutrons and alpha particles. The~coincidence row is for events where two simulated particles interacted with the detector within 20 $\upmu$s, and~so were assigned to a single~event.\label{tab:SoftSimVetoRate}}}
\begin{tabularx}{\textwidth}{CCCC}
\toprule
\textbf{Particle Selection} & \textbf{Initial Rate} & \textbf{Post-Veto Rate} & \textcolor{black}{\textbf{\% Vetoed}}\\
\textbf{-- }& \boldmath{$\mathrm{Events/s}$} &\boldmath{ $\mathrm{Events/s}$} & \textcolor{black}{\textbf{--}}\\
\midrule
Measured Total  & 145.6 & 18.2 &  \textcolor{black}{87.5} \\
Simulated Total & 121.2 & 13.4 &  \textcolor{black}{88.9} \\\midrule
$\gamma$-rays   &  18.1 &  9.2 &  \textcolor{black}{49.4} \\
e$^+$           &   9.4 &  0.8 &  \textcolor{black}{91.0} \\
e$^-$           &   8.7 &  0.8 &  \textcolor{black}{91.0} \\
p$^+$           &  36.4 &  0.2 &  \textcolor{black}{99.3} \\
n$^0$           &   3.8 &  1.2 &  \textcolor{black}{69.5} \\
$\mu^+$         &   0.1 &  0.0 & \textcolor{black}{100.0} \\
$\mu^-$         &   0.1 &  0.0 & \textcolor{black}{100.0} \\
$\alpha$        &  21.9 &  0.1 &  \textcolor{black}{99.7} \\
Activation      &   0.2 &  0.2 &   \textcolor{black}{1.3} \\
Coincidence     &  22.4 &  1.0 &  \textcolor{black}{95.6} \\
\bottomrule
\end{tabularx}

\end{table}
\unskip

\unskip

\unskip

\unskip

\subsection{Event Reconstruction and Spectra}\label{sec:reconstruction}

The events that pass the ACD veto stage are reconstructed with Revan and sorted into PH, CO, MU, and~PA events. Some events are outright rejected for having no valid Compton scattering sequence, among~other reasons. Approximately half of the events pass reconstruction. Table~\ref{tab:SoftSimReconRate} shows the event rates for the measured real data and simulations divided into the reconstructed event type. The~rows are similar to Table~\ref{tab:SoftSimVetoRate}, except~that the muon rows have been removed, because~$100\%$ of the muons were vetoed in the simulation. The~reconstructed event rates underscore that the final product of our data analysis and reconstruction pipeline is predominantly $\gamma$-rays as $67\%$ of the simulated events at this stage are photons, and~$82\%$ of the reconstructed PA events are photons. The~real measured data has an $15\%$ higher event rate than the simulation, and~about twice the MU rate as the simulation. This difference is likely due to discrepancies between the simulation mass model and the~instrument.

\begin{table}[H] 
\caption{A comparison of the reconstruction rates between the measured real data and the simulated data. \textcolor{black}{The table presents the total measured and simulated data, the~simulations by particle type, and~simulated events where 2 particles interacted with ComPair within 20 $\upmu$s during the simulation.} These rates highlight both the effectiveness of the charge particle rejection, that reconstruction maintains $\gamma$-rays dominating over charged particles in our dataset, and~that the reconstructed charged particle interactions are often reconstructed as CO events, while $\gamma$-rays are sometimes classified as MU~events.\label{tab:SoftSimReconRate}}
\begin{tabularx}{\textwidth}{ccCCCC}
\toprule
\textbf{Particle Selection} & \textbf{Recon. Rate} & \textbf{PH Rate} & \textbf{CO Rate} & \textbf{MU Rate} & \textbf{PA Rate}\\
\textbf{--} & \boldmath{$\mathrm{Events/s}$} & \boldmath{$\mathrm{Events/s}$} & \boldmath{$\mathrm{Events/s}$} & \boldmath{$\mathrm{Events/s}$} & \boldmath{$\mathrm{Events/s}$}\\
\midrule
Measured Total & 10.3 & 5.6 & 4.3 & 0.4 & 0.02\\
Simulated Total & 8.9 & 4.7 & 4.1 & 0.2 & 0.02\\\midrule
$\gamma$-rays & 6.0 & 3.2 & 2.7 & 0.09 & 0.02\\
e$^+$ & 0.5 & 0.2 & 0.3 & 0.01 & 0.001\\
e$^-$ & 0.4 & 0.2 & 0.3 & 0.01 & 0.002\\
p$^+$ & 0.1 & 0.05 & 0.06 & 0.03 & 0\\
n$^0$ & 1.0 & 0.5 & 0.4 & 0.03 & 0\\
$\alpha$ & 0.04 & 0.02 & 0.02 & 0.001 & 0\\\midrule
Activation & 0.2 & 0.1 & 0.03 & 0 & 0\\
Coincidence & 0.7 & 0.4 & 0.3 & 0.01 & 0.001\\
\bottomrule
\end{tabularx}

\end{table}

 The reconstructed event rates in Table~\ref{tab:SoftSimReconRate} also highlight a common issue facing MeV telescopes: the charged particle background is misidentified as $\gamma$-ray interactions. In~Table~\ref{tab:SoftSimReconRate}, CO and PA should be purely $\gamma$-ray interactions, but~all non-$\gamma$-ray event types also appear in the simulated CO data, and~$e^-\mathrm{~and~}e^+$ appear in the simulated PA data. For~Revan, it searches first for PH events, because~they only have a single interaction. Then it will search for two tracks and a vertex for PA events, followed by a search for a single track for a MU event. Finally, it will search for a valid Compton scattering sequence and call those CO events. Events without a valid Compton scattering sequence are rejected. This means that charged particles can also end up in the CO category, or~that $\gamma$-rays can be reconstructed as MU events if a tracked Compton electron is misidentified or one of the pair-produced particles escapes. Nevertheless, improvements to the reconstruction methodology is beyond the scope of this work, and~$\gamma$-rays still dominate over the charged particle background \textcolor{black}{following ACD veto and event reconstruction.}

\begin{figure}[H]
    \includegraphics[width=1.0\linewidth]{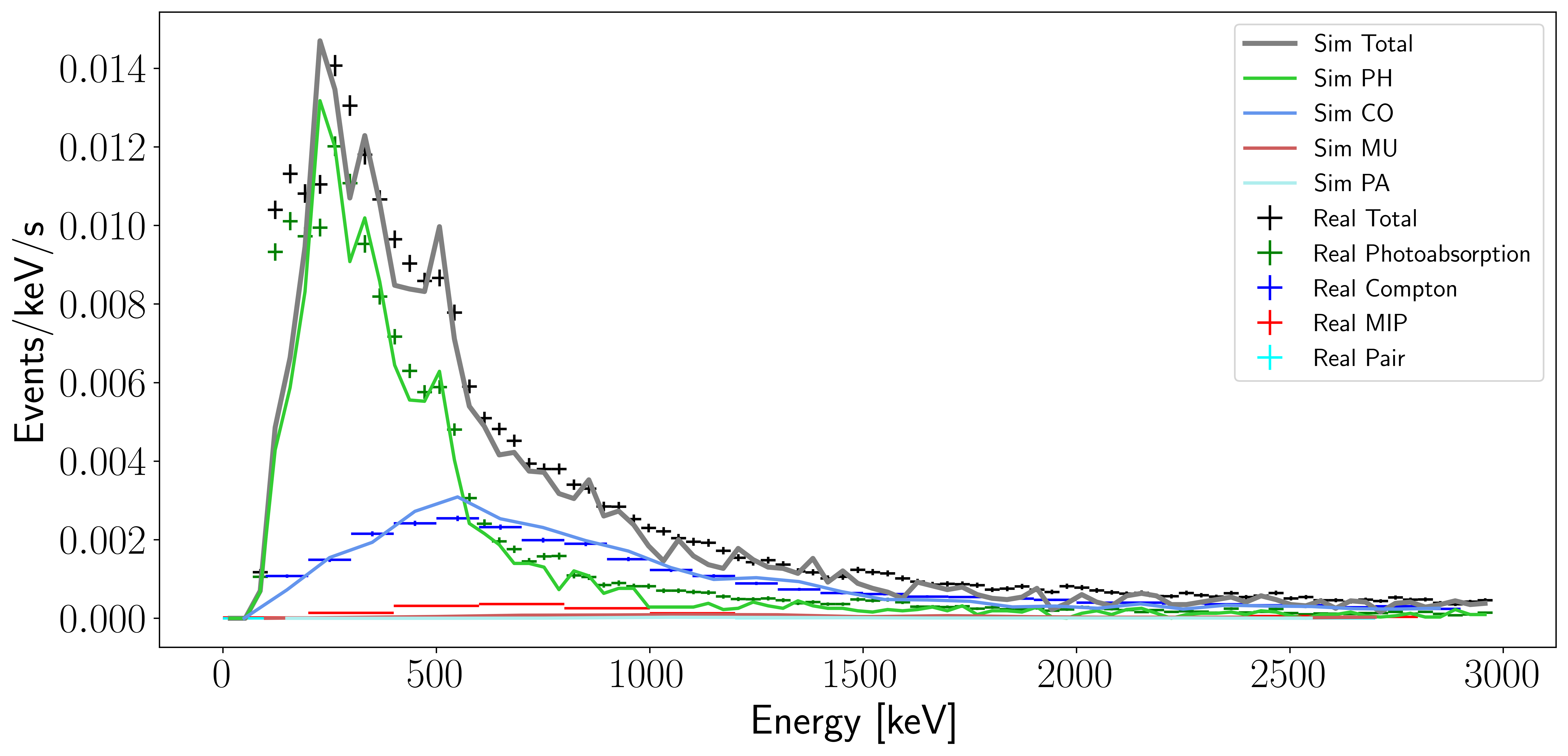}
    \caption{Spectra of the measured energy deposit per event after event reconstruction for the real and simulated data. The~points are the real data with 1$\sigma$ uncertainty within the histogram's bins, and~the lines are the simulations. \textcolor{black}{The uncertainty was estimated assuming Poisson statistics for the number of events within each bin.}}
    \label{fig:SoftSimSpectra}
\end{figure}

Figure~\ref{fig:SoftSimSpectra} shows the spectra of the total energy deposited in the detectors by event for both the real and simulated data, separated by reconstructed event type. The~real data is shown as points with 1 $\sigma$ error bars spanning the histogram bin widths, and~the solid lines are the simulated~data. 

The real and simulated spectra generally agree, except~there seems to be a discrepancy regarding the threshold, which causes the simulation to underestimate the real data by 0.5~Events/s below 200 keV, and~that the real data has a peak in the MU event spectrum near 500 keV that does not appear in the~simulation. 

Also near 500 keV, the~simulated Compton spectrum overestimates the measurement. This could be due to \textcolor{black}{incorrect reconstruction of Compton scattered electrons} as MU tracks when the Compton scattered photon escapes in the real data. In~this case, the~DEE may not fully account for some inefficiencies in the detectors that result in losing the Compton scattered photons. There is also a difference in the prominence of the 511 keV line, which appears sharper in the simulated data than in the measurement. This is likely due to ComPair's energy calibration not accounting for thermal effects, the~temperature change during the flight shifts the gain and energy thresholds of the detectors and \textcolor{black}{smears} the energy response. The~DEE also does not account for the changing temperature, and~therefore leaves a sharp 511 keV line. Despite these two minor discrepancies, the~simulations and the real data are well matched during the balloon~flight.



\subsection{Testing the ACD Hard~Veto}\label{sec:HardVeto}

Looking ahead to AMEGO, possible future challenges are high event rates that severely limit the live time of the instrument and necessitate down linking events due to background particles. With~this in mind, we used the final 11 min of the balloon flight to test a hard ACD veto (the yellow band of Figure~\ref{fig:rates_and_altitude}), where the Trigger Module does not provide an event ID if it receives a primitive from the ACD during data collection. Using a hard veto can be risky, due to the possibility that no event IDs would be produced for event alignment between the other three detectors if the Trigger Module timing is misaligned, hence losing some data. The~goal of this test is to determine that the hard ACD veto did not result in a change in the measured~spectrum.

\begin{figure}[H]
    \subfloat[ \centering]{\includegraphics[height=0.35\textwidth]{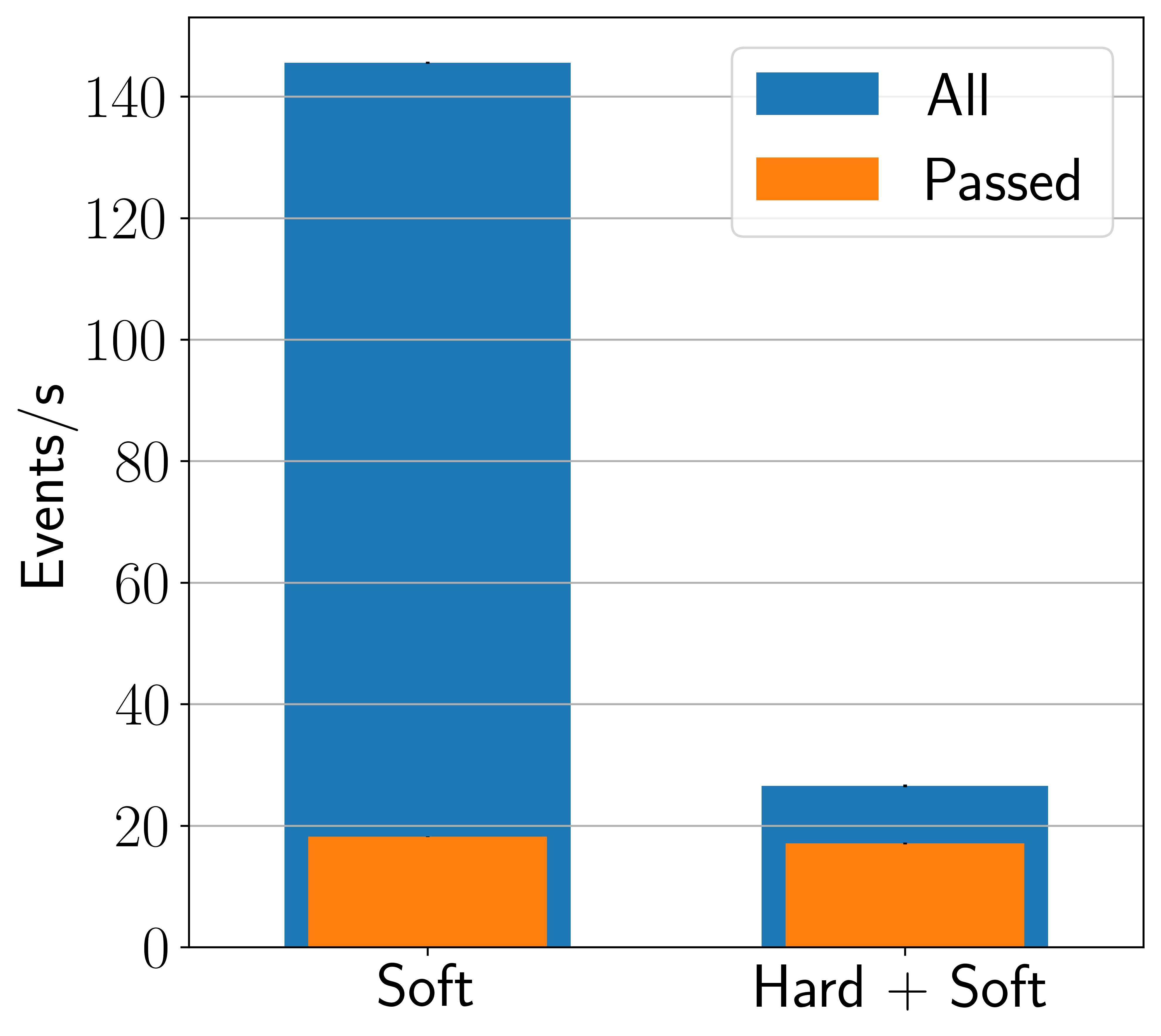}\label{fig:SoftHardRates}}\hskip1ex
    \subfloat[ \centering]{\includegraphics[height=0.35\textwidth]{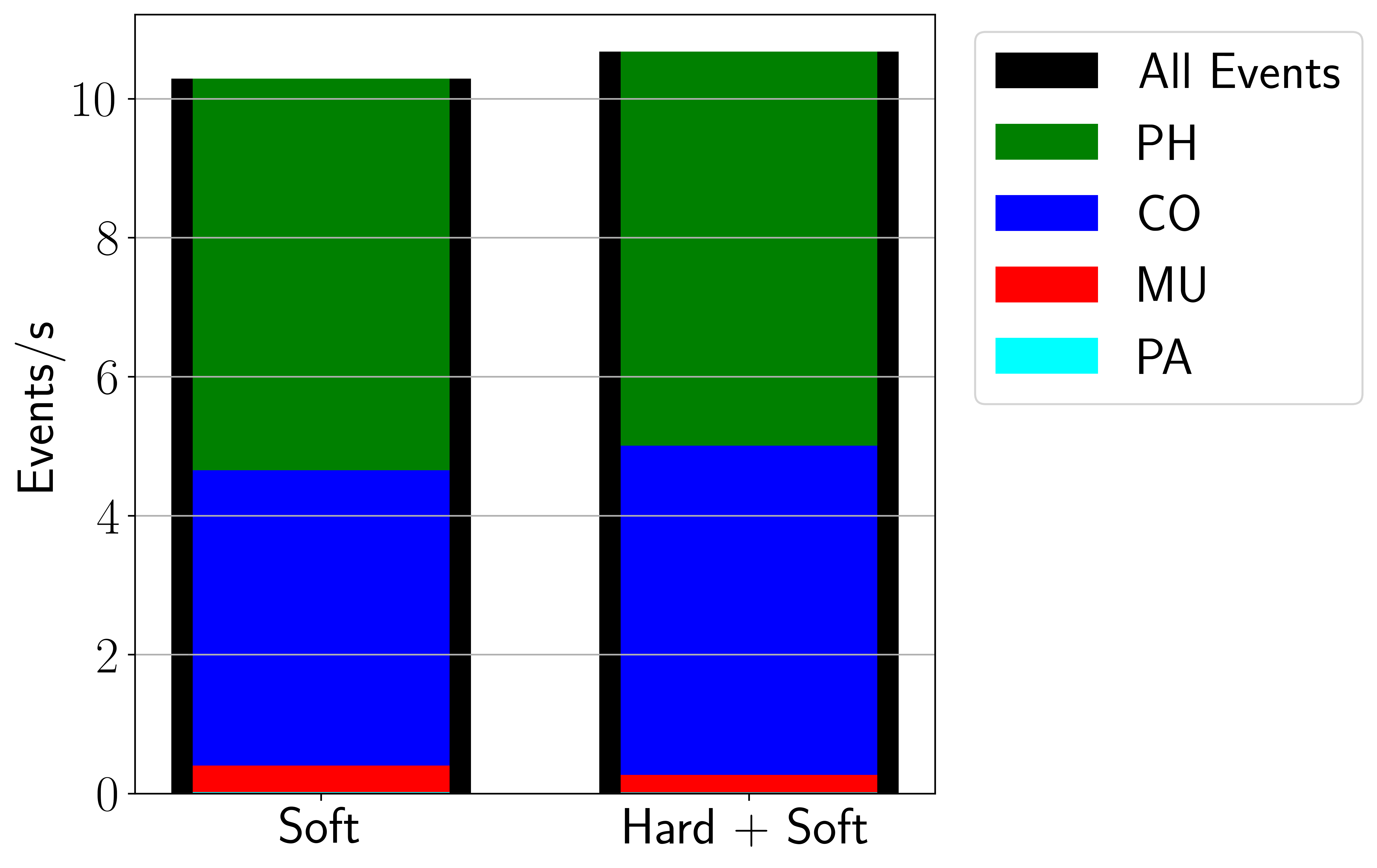}\label{fig:SoftHardRecon}}
    \caption{(\textbf{a}) A comparison of the pre- and post-ACD soft vetos for the datasets collected during the green and yellow bands from Figure~\ref{fig:rates_and_altitude}. The~first column represents the green band and the second column is the yellow band. The~blue and orange bars show the event rates before and after applying a soft ACD veto, respectively. A~concern before this test was that the hard ACD veto might reduce the event rate to below that of the soft veto if Trigger Module timing was off. This result gives confidence that the hard ACD veto could be used in future iterations of ComPair and AMEGO. (\textbf{b}) A comparison of the pre- and post-ACD vetos for the datasets collected during the green and yellow bands from Figure~\ref{fig:rates_and_altitude}. The~first column represents the green bands and the second column is the yellow band. The~bars are divided based on reconstructed event type, and~show little difference between using soft veto only or hard~veto.\label{fig10}}
\end{figure}

Comparing the event rates using the hard veto to the soft veto from Section~\ref{sec:SoftVeto}, we found that the hard veto was not 100\% efficient, as~there were still events with event IDs and energy deposited in the ACD. This necessitates applying an additional soft ACD veto following the hard ACD veto. This is shown in Figure~\ref{fig10}a, where the blue and orange bars represent the event rates before and after the ``ACD Veto'' step of Figure~\ref{fig:DEE1}, respectively, and~the ``Soft'' and ``Hard + Soft'' columns represent the green and yellow bands from Figure~\ref{fig:rates_and_altitude}, respectively. The~hard ACD veto was not 100\% efficient, because~the blue bar of the ``Hard + Soft'' column is higher than either orange bar. The~hard ACD veto being less efficient could have been due to an inefficiency in the ACD sending a primitive to the Trigger Module\textcolor{black}{, losing $\sim 35\%$ of the ACD primitives}. This trigger primitive inefficiency has been observed with \textcolor{black}{the Tracker and Calorimeters around $35\%$ of the time as well}.

\begin{figure}[H]
    \includegraphics[width=0.8
    \linewidth]{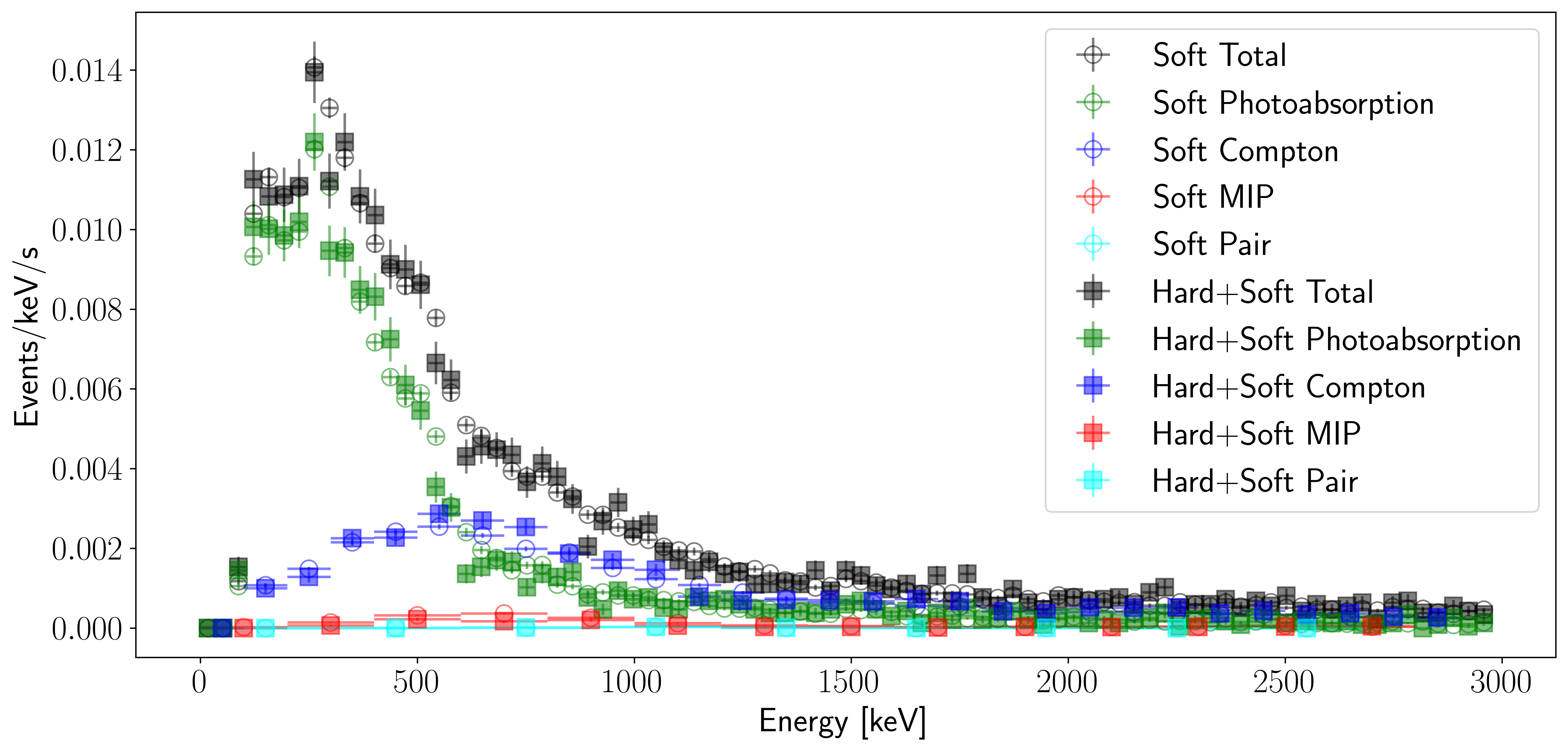}
    \caption{Spectra separated by reconstructed event type for events with soft veto only \textcolor{black}{(squares)} and a hard veto followed by soft veto \textcolor{black}{(circles)}. The~veto scheme did not compromise the measured spectra, validating the hardware~implementation}
    \label{fig:SoftHardSpectra}
\end{figure}

Figure~\ref{fig10}b shows the event rates for the same categories after event reconstruction with Revan. At~this stage, the~``Hard + Soft'' category has retained a higher event rate than the soft-only, which suggests that the event rate variations between the two vetoing schemes are due to statistical fluctuations and not any malfunction of the ACD hard veto. The~spectra in Figure~\ref{fig:SoftHardSpectra} also supports the conclusion that the hard ACD veto did not affect the measured $\gamma$-ray spectrum, as~the soft-veto only and hard-then-soft-veto spectra follow one another closely. For~these reasons, the~hard veto was validated to be used for future iterations of ComPair or AMEGO with lower down link~requirements.

\vspace{-6pt}
\unskip

\section{Conclusions}\label{sec:conclusion}

The ComPair high-altitude balloon flight successfully measured the $\gamma$-ray background at a float altitude of $\approx$40 km above sea level. The~ACD veto and event reconstruction were validated by Monte Carlo simulations of the particle background at float altitudes by comparing event rates and measured spectra. This engineering flight demonstrated the effective rejection of the charged particle background with the ACD and~showed that a hard ACD veto could be used to limit data down link without altering the measured $\gamma$-ray spectrum. Looking forward, the~success of ComPair's 2023 balloon flight sets the stage for ComPair-2~\cite{ComPair2_Caputo_2024} with a $\sim$16 times larger active area, raising the technology readiness levels of several key technologies for AMEGO-X~\cite{AMEGOX_Fleischhack_2021,AMEGOX_Caputo_2022}, a~medium explorer-class concept similar to~AMEGO.

\vspace{6pt} 





\authorcontributions{Conceptualization, D.J.T., B.P., J.M. (Julie McEnery)
, J.E.G., E.H., A.A.M., J.P. and~E.W.; methodology, Z.M., N.C., R.C., C.K., N.K., L.S., M.S., D.S., R.W. and A.A.M.; software, Z.M., S.W., A.Z., N.K., N.C., L.S., L.P., M.S., D.S., S.G. and~P.G.; validation, Z.M.; formal analysis, Z.M., L.S., N.K. and~D.S.; investigation, Z.M., N.C., R.C., C.K., N.K., L.S., L.P., A.J.S., R.W., D.S., A.W.C., T.J.C., J.V., E.K., S.G., I.L.-I., J.M. (John Mitchell), C.S. and~A.A.M.; resources, J.F., A.B., G.A.C. and~S.H.; data curation, Z.M., L.S., N.K. and~D.S.; writing---original draft preparation, Z.M.; writing---review and editing, Z.M., N.K., L.S., N.C., R.C., C.K., M.S., D.S., R.W. and~A.A.M.; visualization, Z.M. and N.K.; supervision, R.C., C.K., R.W. and~A.A.M.; project administration, R.C. and C.K.; funding acquisition, J.M. (Julie McEnery), J.E.G., C.K. and~R.W. All authors have read and agreed to the published version of the manuscript.
}

\funding{\textls[-25]{This work is supported under NASA Astrophysics Research and Analysis (APRA) grants NNH14ZDA001N-APRA, NNH15ZDA001N-APRA, NNH18ZDA001N-APRA, NNH21ZDA001N-APRA.}}

\dataavailability{The raw data supporting the conclusions of this article will be made available by the authors on request.} 

\acknowledgments{The material is based upon work supported by NASA under award number 80GSFC24M0006. Daniel Shy is supported
by the U.S. Naval Research Laboratory’s Jerome and Isabella Karle Distinguished Scholar Fellowship Program. A. W. Crosier and T. Caligiure would like to acknowledge the Office of Naval Research NREIP~Program.}

\conflictsofinterest{Author Emily Kong was employed by the company Technology Service Corporation. The remaining authors declare that the research was conducted in the absence of any commercial or financial relationships that could be construed as a potential conflict of interest. 
} 


  
\abbreviations{Abbreviations}{
The following abbreviations are used in this manuscript:
\\

\vspace{-3pt}
\noindent 
\begin{tabular}{@{}ll}
ACD & Anti-Coincidence Detector\\
AMEGO & All-sky Medium Energy Gamma-ray Observatory\\
AMEGO-X & All-sky Medium Energy Gamma-ray Observatory eXplorer\\
CO & Compton scattering\\
CsI & Cesium Iodide\\
CZT & Cadmium Zinc Telluride\\
DEE & Detector Effects Engine\\
EXPACS & EXcel-based Program for calculating Atmospheric Cosmic-ray Spectrum \\
GEANT4 & GEometry ANd Tracking 4\\
MU & Muon track\\
PA & Pair production\\
PH & Photoabsorption\\
\end{tabular}
}



\begin{adjustwidth}{-\extralength}{0cm}

\reftitle{References}

\PublishersNote{}
\end{adjustwidth}
\end{document}